# Emergent Magnetic Field and Nonzero Gyrovector of the Toroidal Magnetic Hopfion

*Dariia Popadiuk, Elena Tartakovskaya, Maciej Krawczyk, and Kostyantyn Guslienko\**

Magnetic hopfions are localized magnetic solitons with a nonzero 3D topological charge (Hopf index). Herein, an analytical calculation of the magnetic hopfion gyrovector is presented and it is shown that it does not vanish even in an infinite sample. The calculation method is based on the concept of the emergent magnetic field. The particular case of the simplest nontrivial toroidal hopfion with the Hopf index $|Q_H| = 1$ in the cylindrical magnetic dot is considered and dependencies of the gyrovector components on the dot sizes are calculated. Nonzero hopfion gyrovector is important in any description of the hopfion dynamics within the collective coordinate Thiele's approach.

## 1. Introduction

Topological 3D solitons with nonzero Hopf index, named hopfions, were first considered in classical field theories.[1–4] They attracted interest of researchers because such solitons are stable solutions of the Faddeev–Skyrme's Lagrangian. In his pioneering work,[1] Faddeev emphasized that the study of hopfions could become relevant in several theoretical and experimental situations. In fact, the field of study of hopfions turned out to be even wider than Faddeev assumed, covering condensed matter,[5] liquid crystals,[6] photonics,[7] optics,[8] electromagnetism, and gravitation.[9] Recently, stability of the hopfions was studied numerically in chiral ferromagnets[10,11] and in confined ferroelectric nanoparticles.[12] Based on the theoretical predictions,[10] the first experimental observation of magnetic hopfions was carried out in the Ir/Co/Pt multilayer systems.[13] The study of the spin excitation spectra of magnetic hopfions was performed in the recent articles.[14,15]

In addition to considering the hopfions as extremely curious mathematical and physical objects with a nontrivial topology, the question arose of their practical application in data storage and processing devices. In particular, what kind of the magnetic solitons, skyrmions, or hopfions is preferable for information storage based on the magnetic racetrack memory and is it possible to eliminate, in the case of hopfions, the undesirable skyrmion Hall effect?[16–18] As it is well known, the Hall effect (existence of a gyroforce perpendicular to the soliton velocity) for such solitons is attributed to a gyrovector, which, in the case of the magnetic vortices and skyrmions, is not equal to zero and determines their low-frequency dynamics in both 3D and 2D case.[19,20] Therefore, the question of the magnitude of gyrovector of the magnetic hopfion became of practical importance. The components of the emergent magnetic field in the cylindrical coordinates, defining the gyrovector **G** of the magnetic hopfion, were presented in ref. [21]. It was proved unambiguously that the gyrovector axial component $G_z$ of an axially symmetric magnetic hopfion with the Hopf index $|Q_H| = 1$ is equal to zero. As for the zero values of the other two components of the gyrovector, only a plausible assumption was made that they vanish due to the hopfion and system symmetry. Nevertheless, this assumption was accepted on faith in a number of subsequent articles, where the equality to zero of all components of the hopfion gyrovector is mentioned as a proven fact.[22,23] Meanwhile, this issue needs to be clarified and proved because the possible nonzero values of the gyrovector components affect both the three-dimension current-induced hopfion translational motion[24] and the driven by external magnetic field dynamics of spin waves over the background of the hopfion.[22]

In this article, we present a direct analytical calculation of the toroidal hopfion's gyrovector, proving that it does not vanish as a whole and showing that its components are nonzero in different

D. Popadiuk, E. Tartakovskaya, M. Krawczyk
Institute of Spintronics and Quantum Information
Faculty of Physics
Adam Mickiewicz University
61-712 Poznań, Poland

D. Popadiuk, E. Tartakovskaya
Institute of Magnetism
National Academy of Sciences of Ukraine
03142 Kyiv, Ukraine

K. Guslienko
Depto. Polímeros y Materiales Avanzados: Física Química y Tecnología
Universidad del País Vasco
UPV/EHU, 20018 San Sebastián, Spain
E-mail: kostyantyn.gusliyenko@ehu.eus

K. Guslienko
EHU Quantum Center
University of the Basque Country
UPV/EHU, 48940 Leioa, Spain

K. Guslienko
IKERBASQUE
The Basque Foundation for Science
48009 Bilbao, Spain

The ORCID identification number(s) for the author(s) of this article can be found under https://doi.org/10.1002/pssr.202300131.

systems of the coordinates, corresponding to the hopfion and magnetic sample symmetry. This result is especially important for choosing the geometry of driving forces, applied to hopfion, to avoid nondesirable skyrmion Hall effect in the course of the hopfion motion in restricted geometry. It is also useful for understanding the interaction between the magnetic hopfions and elementary excitations (spin waves) on their background.

## 2. Results and Discussion

We start from the general definition of the components of the emergent magnetic field (gyrocoupling density) **B** resulting from an inhomogeneous spin texture $\mathbf{m}(\mathbf{r})$[25]

$$B_i = \frac{1}{2}\varepsilon_{ijk}\mathbf{m}\cdot(\partial_j\mathbf{m}\times\partial_k\mathbf{m}) \quad (1)$$

where $\mathbf{m}(\mathbf{r}) = \mathbf{M}(\mathbf{r})/M_s$ is the unit magnetization vector, $M_s$ is the saturation magnetization, $\partial_j = \partial/\partial x_j$ denote spatial derivatives, and the indices $i, j, k$ correspond to the components of 3D radius vector **r** in an orthogonal coordinate system.

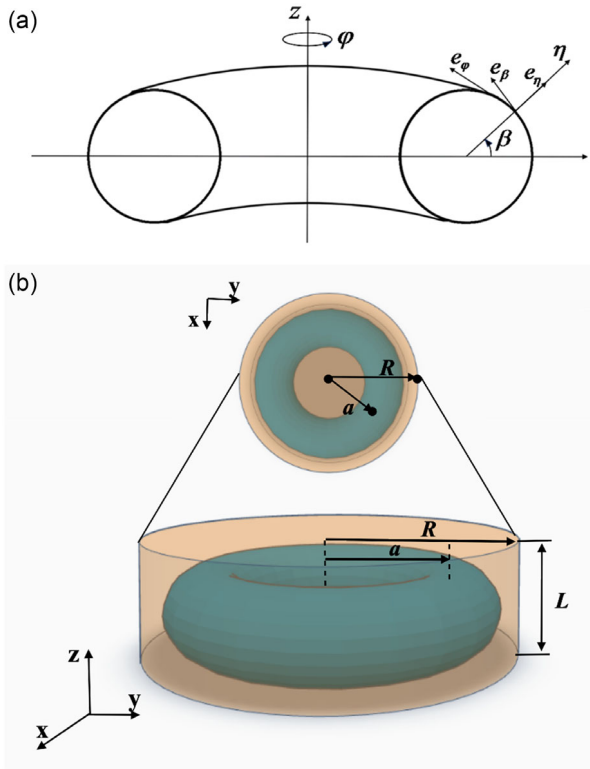

**Figure 1.** a) The toroidal coordinate system $r(\eta, \beta, \varphi)$. The toroidal parameter $\eta$ varies from 0 to $\infty$, the poloidal angle $\beta$ varies from $-\pi$ to $\pi$, and the azimuthal angle $\varphi$ varies from 0 to $2\pi$. The symbols $\mathbf{e}_\eta$, $\mathbf{e}_\beta$, $\mathbf{e}_\varphi$ mark the orthogonal unit vectors along the toroidal axes $(\eta, \beta, \varphi)$, respectively. b) Sketch of a toroidal hopfion with the radius $a$ in the magnetic cylindrical dot with the thickness $L$ and radius $R$. The green torus inside the dot is the isosurface of the constant magnetization component $m_z$ (or constant toroidal parameter $\eta$).

According to the definition of the emergent field by Equation (1), the vector **B** components are expressed in the same orthogonal coordinate system as the radius vector **r** components. The axially symmetric hopfions[2,4] are naturally described in the toroidal coordinates $\mathbf{r}(\eta, \beta, \varphi)$ (see **Figure 1**). The toroidal parameter $\eta$ varies from 0 to $\infty$, the poloidal angle $\beta$ varies from $-\pi$ to $\pi$, and the azimuthal angle $\varphi$ varies from 0 to $2\pi$. The toroidal components of the emergent field **B** of the hopfion are explicitly given by Gladikowski et al.[2]

$$B_\eta = \frac{2N\tau^2}{a^2\sinh(\eta)}\frac{\partial w}{\partial \beta}, \quad B_\beta = -\frac{2N\tau^2}{a^2\sinh(\eta)}\frac{\partial w}{\partial \eta}, \quad B_\varphi = \frac{2M\tau^2}{a^2}\frac{\partial w}{\partial \eta} \quad (2)$$

where $\tau = \cosh(\eta) - \cos(\beta)$, $M, N$ are integer numbers (vorticities in the poloidal and azimuthal directions), $M = \pm 1, \pm 2, \ldots$, $N = \pm 1, \pm 2, \ldots$, $a$ is the hopfion radius, and $w(\eta, \beta) = m_z$ is the out-of-plane magnetization component. The vorticities $M$ and $N$ are defined by the expression $m_x + im_y \propto \exp[i(N\varphi + M\beta)]$. The function $w(\eta, \beta)$ satisfies to the boundary conditions $w(0, \beta) = 1$, $w(\infty, \beta) = -1$. We note that the emergent magnetic field components in Equation (2) are twice bigger than ones defined by Equation (1). We checked equations for the emergent magnetic field by Gladikowski et al.[2] and found that an additional multiplier $(-1/2)$ is necessary to introduce in Equation (2) due to other definition of the emergent magnetic field in ref. [2] in comparison with Equation (1).

The connection of the emergent magnetic field (1) with a hopfion gyrovector is not trivial. If we define the 3D vector $\mathbf{X}(t)$ as the time-dependent magnetic soliton (hopfion) center position, then the dynamical magnetization within the Thiele approach can be written as $\mathbf{m}(\mathbf{r}, t) = \mathbf{m}(\mathbf{r}, \mathbf{X}(t))$. The gyrovector components $G_\alpha$ defined as the volume integrals

$$G_\alpha(\mathbf{X}) = \frac{1}{2}\varepsilon_{\alpha\mu\nu}\int dV \mathbf{m}\cdot\left(\frac{\partial \mathbf{m}}{\partial X_\mu}\times\frac{\partial \mathbf{m}}{\partial X_\nu}\right) \quad (3)$$

determine the gyroforce $\mathbf{F}_G = \mathbf{G}(\mathbf{X})\times d\mathbf{X}/dt$ in the Thiele equation of motion of $\mathbf{X}(t)$.

The Thiele equation of motion can be linearized with respect to small displacement of the soliton with respect to an equilibrium position at $\mathbf{X} = 0$ and one needs to calculate only $\mathbf{G}(\mathbf{X} = 0)$. However, even $\mathbf{G}(0)$ depends on the particular model of the moving soliton, i.e., on the function $\mathbf{m}(\mathbf{r}, \mathbf{X}(t))$. This function is unknown, and several particular trial forms were suggested (for instance, two vortex model[26]) and used for calculation of the soliton gyromodes. The simplest form of the function $\mathbf{m}(\mathbf{r}, \mathbf{X}(t))$ is a rigid soliton ansatz, $\mathbf{m}(\mathbf{r}, \mathbf{X}(t)) = \mathbf{m}_0(\mathbf{r} - \mathbf{X}(t))$, where $\mathbf{m}_0(\mathbf{r})$ is a static soliton profile at $\mathbf{X} = 0$. The rigid soliton ansatz is, strictly speaking, applicable only for infinite media, where translation invariance with respect to the arbitrary soliton displacement $\mathbf{X}$ is valid. We use below the rigid motion ansatz to express the hopfion gyrovector via the emergent magnetic field (1). Accounting the relations $\partial/\partial x_j = -\partial/\partial X_j$, the hopfion gyrovector components can be written as volume integrals

$$G_\alpha(0) = \int dV B_\alpha \quad (4)$$

from the emergent magnetic field components $B_\alpha$. We use below the field components in the toroidal coordinates (2) or in the cylindrical coordinates $(\rho, \varphi, z)$ assuming volume integration over a cylindrical magnetic dot.

We choose the function $w(\eta, \beta)$ in the form $w(\eta, \beta) = 1 - 2\tanh^2(\eta)$ suggested by Hietarinta et al.[3,4] for the case of the simplest hopfion with $M = 1$, $N = 1$ and use the components of the radius vector $\mathbf{r}$ in the cylindrical coordinates, $\mathbf{r}(\rho, \varphi, z)$. This form follows from the explicit expression

$$m_z = 1 - \frac{8\rho^2 a^2}{(a^2 + \rho^2 + z^2)^2} \quad (5)$$

given in the article by Hietarinta et al.[3] for the Hopf index $|Q_H| = 1$. Such magnetic hopfions are of the main interest because the hopfion energy increases with the increase of the index $|Q_H|$.[2–4] The isosurfaces of the constant magnetization component $m_z$ (or constant toroidal parameter $\eta$) are nonintersecting tori described by the equation $z^2 + (\rho - a\coth(\eta))^2 = a^2/\sinh^2(\eta)$. The torus centers are located at the rings $\rho = a\coth(\eta)$ in the $xy$-plane and the torus radii are equal to $a/\sinh(\eta)$. The magnetization component is $m_z = +1$ in the hopfion center at $\mathbf{r} = 0$ and $m_z = -1$ at the ring $\rho = a$ in the limit $\eta \to \infty$ that justifies using the parameter $a$ as the hopfion radius. The magnetic hopfion radius is analogues to the domain wall width in ferromagnets.

For the particular case of the axially symmetric hopfions with the Hopf index $Q_H = \pm 1$ considered in refs. [21–24] and using the relation $Q_H = MN$, we chose the vorticities $N = 1$, $M = 1$. Then, using the volume element in the toroidal coordinates $dV = a^3 \sinh(\eta) d\eta d\beta d\varphi/\tau^3$ and conducting the volume integration assuming infinite media, we get the gyrovector of a finite magnitude ($|\mathbf{G}| \neq 0$) in the form

$$\mathbf{G} = (G_\eta, G_\beta, G_\varphi) = 8\pi^2 a^2 (0, -\pi/2, 1) \quad (6)$$

The hopfion gyrovector $\mathbf{G} = (G_\eta, G_\beta, G_\varphi)$ (6) has nonzero $\beta$ and $\varphi$ components and corresponding nonzero length. This is in contrast to ref. [21], where zero global gyrovector of the toroidal hopfion was calculated for an infinite media. The gyrovector is an important parameter to describe the hopfion dynamics within the Thiele approach. Therefore, this discrepancy should be clarified.

Let us calculate the global gyrovector $G_\alpha = \int dV B_\alpha$ components for a finite magnetic system (a cylindrical magnetic dot, see Figure 1b). We use the angular parameterization for the dot magnetization components via spherical angles $\Theta, \Phi$: $m_z = \cos\Theta$, $m_x + im_y = \sin\Theta \exp(i\Phi)$, and the cylindrical coordinates $\mathbf{r}(\rho, \varphi, z)$ for the radius vector $\mathbf{r}$ and the cylindrical components of the field $\mathbf{B}$, which fit to the symmetry of the dot. The magnetization spherical angles are functions of the radius vector, $\Theta = \Theta(\mathbf{r})$, $\Phi = \Phi(\mathbf{r})$. Following the theory of 2D magnetic solitons (vortices and skyrmions), it is natural to choose the hopfion magnetization spherical angles in axially symmetric form, $\Theta(\mathbf{r}) = \Theta(\rho, z)$, $\Phi(\mathbf{r}) = N\varphi + \gamma(\rho, z) + \Phi_0$, where the constant phase $\Phi_0$ distinguishes the Bloch and Neel hopfions. The variable soliton helicity $\gamma(\rho, z)$ is of principal importance to secure nonzero Hopf index and nonzero gyrovector of the 3D magnetization textures, such as a toroidal hopfion[21] or Bloch point hopfion,[27] localized and nonlocalized magnetic topological solitons, respectively.

Substituting the magnetization spherical angles $\Theta(\mathbf{r})$, $\Phi(\mathbf{r})$ to (Equation (1)), one can get the components of the emergent field in the cylindrical coordinates

$$B_\rho = -N\frac{\sin\Theta}{\rho}\frac{\partial \Theta}{\partial z}, \quad B_\varphi = \sin\Theta\left(\frac{\partial \Theta}{\partial z}\frac{\partial \gamma}{\partial \rho} - \frac{\partial \Theta}{\partial \rho}\frac{\partial \gamma}{\partial z}\right),$$
$$B_z = N\frac{\sin\Theta}{\rho}\frac{\partial \Theta}{\partial \rho} \quad (7)$$

The emergent magnetic field components given by Equation (7) depend only on the derivatives $\partial \gamma/\partial \rho$, $\partial \gamma/\partial z$. Therefore, there is no difference between the Bloch and Neel hopfions. The Bloch and Neel hopfions have the same emergent magnetic field configurations and the same gyrovectors. The $\rho$ and $z$ components of the emergent magnetic field are calculated substituting to Equation (7) the hopfion magnetization $m_z = \cos\Theta$ given by Equation (5)

$$B_\rho(\rho, z) = \frac{32a^2\rho z}{(a^2 + \rho^2 + z^2)^3}, \quad B_z(\rho, z) = \frac{16a^2(a^2 + z^2 - \rho^2)}{(a^2 + \rho^2 + z^2)^3} \quad (8)$$

Both $\rho$ and $z$ components of the emergent magnetic field are determined by the $z$ component of the magnetization, $m_z(\rho, z)$. It follows from the expressions (Equation (7)) that the $\rho$ component of the hopfion global gyrovector (Equation (4)) is equal to zero, $G_\rho = 0$, for any magnetization distribution, where the component $m_z(\rho, z)$ is even function of the coordinate $z$ and for any sample shape which satisfies the symmetry with respect to change of $z$ sign, $z \to -z$. The class of such magnetization configurations includes the toroidal hopfions and Bloch points. The $z$ component of the gyrovector is also equal to zero, $G_z = 0$, for the toroidal hopfions in infinite media according to the equation for $m_z(\rho, z)$ by Hietarinta et al.[3] For a finite cylindrical ferromagnetic dot with thickness $L$ and radius $R$, the component $G_z(L, R)$ decreases as $a/R$ at $R \to \infty$ keeping the dot aspect ratio $L/R = \text{const}$. This decrease is a consequence of the fact that the toroidal hopfion is a localized topological soliton, i.e., $\mathbf{m}(\mathbf{r}) \to \mathbf{m}_0(0, 0, 1)$ at $|\mathbf{r}|/a \gg 1$. The toroidal hopfion is schematically shown in Figure 1b. The function $G_z(L, R)$ calculated by using the volume integration of the expression for $B_z(\rho, z)$ in Equation (8) using Equation (5) has the form

$$G_z(L, R) = 16\pi a \frac{(R/a)^2}{(1 + (R/a)^2)^{3/2}}\left(\frac{t}{1 + t^2} + \arctan(t)\right),$$
$$t = \frac{L/a}{2\sqrt{1 + (R/a)^2}} \quad (9)$$

where $\beta = L/R$ is the aspect ratio of the cylindrical dot. The dependence of $G_z(L, R)$ on the dot parameters is shown in **Figure 2**.

The calculation of the in-plane $G_\varphi$ component of the hopfion gyrovector is more complicated because it involves the hopfion helicity $\gamma(\rho, z)$. There is the connection between the cylindrical $(\rho, \varphi, z)$ and toroidal $(\eta, \beta, \varphi)$ coordinates[28]

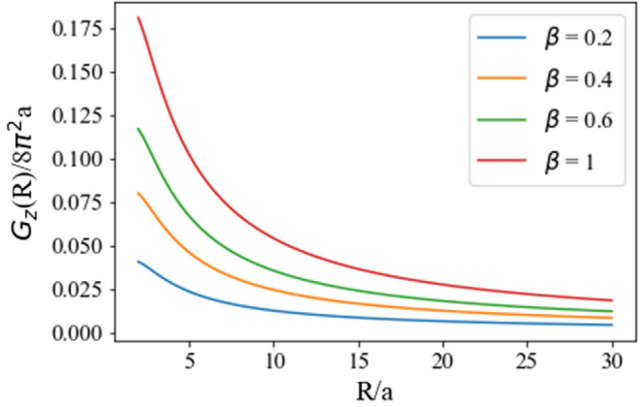

**Figure 2.** Dependence of the out-of-plane hopfion gyrovector component $G_z$ on the dot radius $R$ for different dot aspect ratios, $\beta = L/R$.

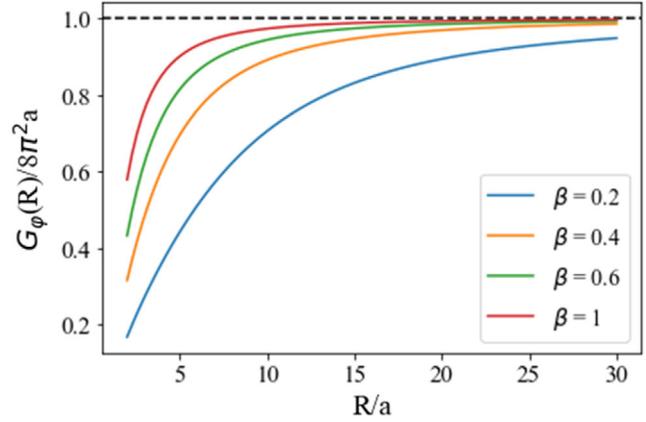

**Figure 3.** Dependence of the hopfion gyrovector in-plane component $G_\varphi$ on the dot radius $R$ for different dot aspect ratios, $\beta = L/R$.

$$\rho = a\frac{\sinh\eta}{\tau}, \quad z = a\frac{\sin\beta}{\tau}, \quad \varphi = \varphi \tag{10}$$

This connection can be rewritten in the compact form

$$z + i\rho = ia\coth\frac{1}{2}(\eta + i\beta), \quad \eta + i\beta = \ln\frac{z + i(\rho + a)}{z + i(\rho - a)} \tag{11a}$$

The definition of the magnetization azimuthal angle $\Phi$ in the toroidal coordinates according to refs. [2,4] is $\Phi(\beta,\varphi) = N\varphi + M\beta + \text{const}$. Therefore, the hopfion helicity $\gamma$ for $M = 1$ can be immediately represented via the toroidal angle $\beta$, $\gamma(\rho,z) = \beta(\rho,z) + \text{const}$, i.e., the helicity has a pure geometrical origin. The explicit form of the function $\beta(\rho,z)$ is given by the expression

$$\beta(\rho,z) = Im\left[\ln\frac{z + i(\rho + a)}{z + i(\rho - a)}\right] \tag{11b}$$

Substituting the function $\gamma(\rho,z)$ to the definition of the $\varphi$ component of the emergent magnetic field, $B_\varphi$, we get the expression

$$B_\varphi(\rho,z) = \frac{32a^3\rho}{(a^2 + \rho^2 + z^2)^3} \tag{12}$$

The emergent magnetic field length $|\mathbf{B}(\mathbf{r})|^2 = (4a)^4/(a^2 + \rho^2 + z^2)^4 = 16\tau^4/a^4\cosh^4(\eta)$, where $\mathbf{B}$ components are defined by Equation (8) and (12), is conserved under transformation from the toroidal $(\eta,\beta,\varphi)$ to the cylindrical $(\rho,\varphi,z)$ coordinates.

Conducting integration of the expression for $B_\varphi(\rho,z)$ over the cylindrical dot volume, we get that the in-plane $\varphi$ component of the global gyrovector, $G_\varphi$, is not equal to zero even for an infinite sample. The function $G_\varphi(L,R)$ has the form

$$G_\varphi(\beta, R) = 64\pi a\left(\frac{\beta R/a}{\sqrt{1 + (\beta R)^2/(4a^2)}}F\left(\frac{R/a}{\sqrt{1 + (\beta R)^2/(4a^2)}}\right)\right.$$
$$\left. + \frac{2(R/a)^3}{(1 + (R/a)^2)^{3/2}}F\left(\frac{\beta R/a}{2\sqrt{1 + (R/a)^2}}\right)\right) \tag{13}$$

where $F(t) = -\frac{t}{4(1+t^2)^2} + \frac{t}{8(1+t^2)} + \arctan(t)/8$, and $\beta = L/R$.

The function $G_\varphi(\beta, R)$ increases monotonically with the dot radius $R$ increasing at the fixed $L/R$ dot aspect ratio and saturates at the value $G_\varphi(L/R, R \to \infty) = 8\pi^2 a$ (in units of the angular momentum density $M_s/\gamma$) (see **Figure 3**). The function $B_\varphi(\rho,z)$ has definite sign (positive for the given hopfion with the azimuthal and poloidal vorticities $N = 1, M = 1$) and therefore the volume integral $G_\varphi = \int dV B_\varphi$ is not equal to zero for any sample volume. The calculated hopfion gyrovector for the cylindrical dot has the form $\mathbf{G} = (0, G_\varphi, G_z)$.

We note that it is evident from Equation (8) and (12) expressed in the cylindrical coordinates that the magnetic emergent field of the toroidal hopfion is solenoidal, $\text{div}\mathbf{B} = 0$, as it should be accounting the relation $\mathbf{B} = \text{rot}\mathbf{A}$, where $\mathbf{A}$ is the vector potential of the emergent electromagnetic field.[2] There is no necessity to introduce the Dirac string as it was done in ref. [26] for the Bloch point hopfion to satisfy the condition $\text{div}\mathbf{B} = 0$.[26]

Alternative approach to the problem is calculation of the direction cosines for the transformation from the toroidal coordinates to the cylindrical coordinates,[28] and change the toroidal components of the emergent field $(B_\eta, B_\beta, B_\varphi)$ given in the article by Gladikowski et al.[2] to the cylindrical ones $(B_\rho, B_\varphi, B_z)$. Then, the volume integration in Equation (4) can be conducted using the cylindrical coordinates $\mathbf{r}(\rho,\varphi,z)$, which correspond to the symmetry of the system. The component $B_\eta = 0$; therefore, the cylindrical components of the emergent field are

$$B_\rho = \frac{N}{a^2}\tau\sin(\beta)\frac{\partial w}{\partial \eta}, \quad B_\varphi = \frac{M\tau^2}{a^2}\frac{\partial w}{\partial \eta},$$
$$B_z = -\frac{N}{a^2}\frac{\tau}{\sinh(\eta)}(\cosh(\eta)\cos(\beta) - 1)\frac{\partial w}{\partial \eta} \tag{14}$$

where $w(\eta) = m_z(\eta) = 1 - 2\tanh^2(\eta)$.

Substituting the toroidal coordinates $(\eta, \beta, \varphi)$ in Equation (14) to the cylindrical coordinates $(\rho, \varphi, z)$ using the relations of the coordinates given by Equation (10) and (11a,b) after some algebra, one can get the emergent magnetic field components in the form of Equation (8) and (12). Following this approach, it is not necessary to introduce explicitly the hopfion helicity $\gamma(\rho, z)$ used in Equation (7).

We calculated the axially symmetric hopfion gyrovector components $G_\alpha = \int dV B_\alpha$ in the toroidal coordinate system $(\eta, \beta, \varphi)$, which is most physical one because according to refs. [2,4] the magnetization field has the simplest representation in these coordinates reflecting the toroidal hopfion symmetry. Assuming integration over the cylindrical dot volume, we calculated the gyrovector components $G_\alpha$ in the cylindrical coordinate system $\mathbf{r}(\rho, \varphi, z)$, which reflects simultaneously the hopfion symmetry and dot symmetry. Difference of our treatment of the hopfion gyrovector (4) and the global hopfion gyrovector $\tilde{\mathbf{G}}$ in refs. [21,23,29] is due to definition $\tilde{\mathbf{G}} = \int dV \mathbf{B}$. Using the decomposition $\mathbf{B} = (B_\alpha \mathbf{e}_\alpha)$ in some orthogonal unit vector basis $\mathbf{e}_\alpha(\mathbf{r})$, one can calculate the gyrovector components $\tilde{G}_\alpha(\mathbf{r}) = \tilde{\mathbf{G}} \cdot \mathbf{e}_\alpha(\mathbf{r}) = \int dV' B_\beta(\mathbf{r}')(\mathbf{e}_\alpha(\mathbf{r}) \cdot \mathbf{e}_\beta(\mathbf{r}'))$. In general, the dot product $(\mathbf{e}_\alpha(\mathbf{r}) \cdot \mathbf{e}_\beta(\mathbf{r}')) \neq \delta_{\alpha\beta}$, except the case of the Cartesian unit vectors $e_\alpha$. Therefore, the gyrovector components $G_\alpha$ and $\tilde{G}_\alpha$ are not equivalent is some general (curvilinear) coordinate systems. Formally, $\tilde{G}_\alpha = 0$ in the Cartesian coordinates $\alpha = x, y, z$ in infinite media for an axially symmetric hopfion. However, the Cartesian components $\tilde{G}_\alpha$, explicitly or implicitly used in refs. [21,23,29] and resulted in the conclusion about nullification of the global hopfion gyrovector, are unphysical because they do not correspond to the toroidal hopfion symmetry.

We note that the dimensionless hopfion gyrovector components $G_\alpha/a$ shown in Figure 2 and 3 are plotted as function of the dot radius in units of the hopfion radius, $R/a$. The equilibrium value of $a$ should be obtained from the total magnetic energy minimization, and apparently depends on the dot magnetic (the exchange stiffness, Dzyaloshinskii–Moriya interaction, magnetic anisotropy, etc.) and geometrical parameters. Therefore, the calculated dependences of $G_\alpha/a$ on $R/a$ are universal and determined solely by the magnetization configuration $\mathbf{m}(\mathbf{r})$ topology, whereas the hopfion radius $a$ depends on the particular model of the magnetic dot. To calculate the emergent magnetic field (Equation (1)) and gyrovector (Equation (4)) in absolute units, the multiplier $\Phi_0/2\pi$ (where $\Phi_0 = h/2e$ is the flux quantum, $h$ is the Planck constant, $e$ is the electron charge) should be added to Equation (1).

The calculation approach described above is based on the concept of the emergent magnetic field $\mathbf{B}(\mathbf{r})$ introduced by Equation (1) and calculation of the field components and their volume averages for the toroidal hopfion spin texture in the appropriate curvilinear coordinate systems. Although this emergent magnetic field differs from real magnetic field defined in the classical electrodynamics and is "fictious" in some sense, it leads to some experimentally measurable effects. We mention here the topological Hall effect (influence of the emergent magnetic field on the conductivity electrons scattering) and skyrmion Hall effect (deflection of the magnetic soliton motion from the driving force direction due to the gyroforce). A theoretical approach how to use the topological Hall resistance to electrically detect the magnetic hopfion 3D spin textures was recently suggested by Göbel et al.[29]

## 3. Conclusions

We considered the simplest nontrivial toroidal hopfion with the Hopf index $|Q_H| = 1$ in the cylindrical magnetic dot and calculated the dependencies of the hopfion gyrovector components on the dot sizes. We demonstrated by the analytical calculations that the magnetic hopfion gyrovector is not equal to zero and does not vanish even in the limit of an infinite sample. Namely, two components of the gyrovector, $G_z$ and $G_\varphi$, are nonzero. The out-of-plane $z$ component of the gyrovector ($G_z$) goes to zero increasing the dot radius; however, the in-plane $\varphi$-component ($G_\varphi$) remains finite. The calculation method is based on the concept of the emergent magnetic field $\mathbf{B}(\mathbf{r})$, which is expressed via spatial derivatives of the magnetization field $\mathbf{m}(\mathbf{r})$. The calculated components of the hopfion emergent field and gyrovector can be used for calculations of the topological Hall effect and skyrmion Hall effect of the toroidal magnetic hopfions, respectively.

### Acknowledgements

The authors acknowledge support by the Norwegian Financial Mechanism 2014-2021 trough project UMO-2020/37/K/ST3/02450. K.G. acknowledges support by IKERBASQUE (the Basque Foundation for Science). The research of K.G. was funded in part by the Spanish Ministry of Science and Innovation grant PID2019-108075RB-C33/AEI/10.13039/501100011033.

### Conflict of Interest

The authors declare no conflict of interest.

### Data Availability Statement

The data that support the findings of this study are available from the corresponding author upon reasonable request.

### Keywords

hopfions, magnetic dots, magnetization textures

Received: April 5, 2023
Revised: May 10, 2023
Published online: